\documentclass[11pt]{article}
\usepackage{graphicx}  % standard LaTeX graphics tool

\usepackage{a4}
\usepackage{amssymb,latexsym}%,amsmath}
\usepackage[mathscr]{eucal}
\usepackage{citesort}
\usepackage{url}
\usepackage{ntheorem}

\newcommand{\hypkz}{\hyp_{\kappa_0}}
\newcommand{\hnabla}{\hat \nabla}
\newcommand{\hpsi}{\hat \psi}

\newcommand{\repf}{{\hat f}}%
\newcommand{\repe}{{\hat e}}%

\newcommand{\mB}{{\mathrm B}}

\newcommand{\hK}{\hat K}

\newcommand{\TBm}{{\mathrm{ TB}}}

\newcommand{\fourg}{\,{}^{4}g}%

\newcommand{\Nu}{V_u}

\newcommand{\ch}{\bh}%{\check h}
\newcommand{\ourV}{\mathbb V}

\newcommand{\rd}{\,{ d}} % exterior differential

      \newcommand{\bh}{{\breve h}} % standard metric on S^2

      \newcommand{\TB}{Trautman--Bondi\ }

\newcommand{\hst}{{\breve{h}}} % standard round metric on the two
\newcommand{\FS}       %{F_1} %
                  {F}
\newcommand{\HS} %{F_2}
       {H_{\mbox{\scriptsize volume}}}

\newcommand{\zD}{\mathring{D}}%
\newcommand{\zK}{\mathring{K}}%
\newcommand{\zomega}{\mathring{\omega}}%

\newcommand{\ourU}{\mathbb U}%
\newcommand{\eeal}[1]{\label{#1}\end{eqnarray}}
\newcommand{\bed}{\begin{deqarr}}
\newcommand{\eed}{\end{deqarr}}
\newcommand{\bedl}[1]{\begin{deqarr}\label{#1}}
\newcommand{\eedl}[2]{\arrlabel{#1}\label{#2}\end{deqarr}}

\newcommand{\bel}[1]{\begin{equation}\label{#1}}
\newcommand{\bea}{\begin{eqnarray}}
\newcommand{\bean}{\begin{eqnarray}\nonumber}
\newcommand{\bbean}{\begin{eqnarray*}}
\newcommand{\eean}{\end{eqnarray*}}
\newcommand{\beal}[1]{\begin{eqnarray}\label{#1}}
\newcommand{\eea}{\end{eqnarray}}
\newcommand{\nn}{\nonumber}
\newcommand{\Eq}[1]{Equation~\eq{#1}}

\newcommand{\be}{\begin{equation}}
\newcommand{\eeq}{\end{equation}}
\newcommand{\ee}{\end{equation}}
\newcommand{\beqa}{\begin{eqnarray}}
\newcommand{\eeqa}{\end{eqnarray}}
\newcommand{\beqan}{\begin{eqnarray*}}
\newcommand{\eeqan}{\end{eqnarray*}}
\newcommand{\ba}{\begin{array}}
\newcommand{\ea}{\end{array}}

\newcommand{\const}{\mbox{\rm const}} %constants

\newcommand{\hyp}{\mycal S}

\newcommand{\mcM}{{\mycal M}}
\newcommand{\mcD}{{\mycal D}}

\theorembodyfont{\upshape}

\DeclareFontFamily{OT1}{rsfs}{} \DeclareFontShape{OT1}{rsfs}{m}{n}{
<-7> rsfs5 <7-10> rsfs7 <10-> rsfs10}{}
\DeclareMathAlphabet{\mycal}{OT1}{rsfs}{m}{n}
\def\scri{{\mycal I}}%
\def\scrip{\scri^{+}}%

{\catcode `\@=11 \global\let\AddToReset=\@addtoreset}
\AddToReset{equation}{section}

\newcounter{mnotecount}[section]

\renewcommand{\themnotecount}{\thesection.\arabic{mnotecount}}

\newcommand{\mnote}[1]%{}%
{\protect{\stepcounter{mnotecount}}$^{\mbox{\footnotesize
$%\!\!\!\!\!\!\,
\bullet$\themnotecount}}$ \marginpar{%\color{red}%
\raggedright\tiny\em $\!\!\!\!\!\!\,\bullet$\themnotecount: #1} }

\newcommand{\warn}[1]%{}
{\protect{\stepcounter{mnotecount}}$^{\mbox{\footnotesize
$%\!\!\!\!\!\!\,
\bullet$\themnotecount}}$ \marginpar{%\color{red}%
\raggedright\tiny\em $\!\!\!\!\!\!\,\bullet$\themnotecount: {\bf
Warning:} #1} }

\newcommand{\R}{\mathbb R}

\newcommand{\backn}{{}^{n}b}

\newcommand{\eq}[1]{(\ref{#1})}

\newcommand{\tmcM}{\,\,\,\,\,\widetilde{\!\!\!\!\!\mcM}}
\newcommand{\beqar}{\begin{deqarr}}
\newcommand{\eeqar}{\end{deqarr}}

\newcommand{\beaa}{\begin{eqnarray*}}
\newcommand{\eeaa}{\end{eqnarray*}}

\newcommand{\tr}{\mbox{tr}}

\newcommand{\bmetric}{{b}} % metryka tla

\newcommand{\jexp}[1]{{\rm e}^{#1}}
\newcommand{\tlambda}{\tilde \lambda} % extrinsic curvature of
\newcommand{\ccalD}{\;\breve{\!\cal D}}

\newcommand{\tmet}{x^{2}g}%{\tilde\met} % conformally rescaled metric on a
\newcommand{\beq}{\begin{equation}}

\newcommand{\zpsi}{{\mathring \psi}}%
\newcommand{\Scrip}{\scrip}%

\begin{document}
\title{An angular momentum bound at null infinity}

\author{Piotr T.\ Chru\'sciel\thanks{%Currently at the Albert Einstein Institute, Golm.
%Partially supported by a Polish
%Research Committee grant 2 P03B 073 24.
E-mail
    \protect\url{Chrusciel@maths.ox.ac.uk}, URL
    \protect\url{ www.phys.univ-tours.fr/}$\sim$\protect\url{piotr}}
  \\ LMPT,
F\'ed\'eration Denis Poisson, Tours
\\
Mathematical Institute and Hertford College, Oxford\\
%Facult\'e des Sciences\\ Parc
%  de Grandmont\\ F37200
\\
\\
  Paul Tod\thanks{{E--mail} \protect\url{Tod@maths.ox.ac.uk}}
\\
Mathematical Institute and St John's College, Oxford}

\maketitle

\begin{abstract}
We prove an inequality relating the trace of the extrinsic
curvature, the total angular momentum, the centre of mass, and the
Trautman-Bondi mass for a class of gravitational initial data sets
with constant mean curvature extending to null infinity. As an
application we obtain non-existence results for the asymptotic
Dirichlet problem for CMC hypersurfaces in stationary space-times.
\end{abstract}

\section{Introduction}
\label{introduction}

Let $(\hyp,g,K)$ be an $n$-dimensional, $n\ge 3$,   constant mean
curvature (CMC) general relativistic initial data set with
cosmological constant $\Lambda$ (possibly zero), thus,
\beal{c1}
 &
 R = |K|^2 - (\tr_g K)^2 + 2 \Lambda + 16 \pi \rho
 \;,
 &
 \\
 &
 D_i K^{ij}
 =- 8\pi \mu^j\;, \quad  D_i  (\tr_g K)=0
 \;.
 &
 \eeal{c2}
Here $\rho$ is the matter energy density, and $\mu^j$ is the matter
momentum vector.

 There is a   transformation which maps
such initial data sets with $\tr_g K=\kappa$
%and $\Lambda=0$
to new initial data sets $(\hyp, g ,\hK)$ with $\tr_g \hK = 0$ and
$\Lambda $ shifted by $-(n-1)\kappa^2/2n$: Indeed, if
%$
\bel{Ktransf}
 \hK_{ij} = K_{ij}- \frac \kappa n g_{ij}  \;,
\ee
%  $,
then \eq{c2} still holds with $K$ replaced by $\hat K$, while
\eq{c1} becomes
\beal{c3}
 &
 R = |\hK|^2 - \underbrace{(\tr_g \hK)^2}_{0} + 2 \underbrace{(\Lambda - \frac {n-1}{2n}\kappa^2)}_{\hat \Lambda} +
 16 \pi \rho
 \;.
 &
 \eea
\Eq{Ktransf} allows one to go back and forth from CMC hyperboloidal
initial data sets in space-times with $\Lambda=0$  to initial data
sets in asymptotically anti-de Sitter space-times with $\Lambda <0$.

The object of this note is to point out that this transformation,
together with the known bounds on total angular momentum and centre
of mass for asymptotically anti-de Sitter
space-times~\cite{Maerten,CMT}, implies a striking angular-momentum
bound for  CMC hyperboloidal initial data which are asymptotically
flat at null infinity, see \eq{Jin2n} below.

Our analysis complements Dain's recent upper bound on angular
momentum~\cite{Dain:2006} at spatial infinity for axi-symmetric
solutions with two asymptotically flat regions.

As an interesting application, we obtain   non-existence results for
hypersurfaces as above in stationary space-times, see
Section~\ref{Sobs} below.

Before presenting our inequality it is useful to review the
definitions of global charges both with $\Lambda=0$ and $\Lambda
<0$; we start with the latter.

\section{Global charges for asymptotically anti-de Sitter initial data}
\label{SAadS}

For the purposes of this work,  an $n$-dimensional initial data set
$(\hyp, g,K)$ will be called \emph{asymptotically anti-de Sitter}
(adS) if $\hyp$ contains an asymptotic region, diffeomorphic to the
complement of a ball in $\R^n$, in which $K$ asymptotes to zero
while $g$ asymptotes to a Riemannian background metric
\bel{backgmet}
 b=dr^2+
 \sinh^2 (r)\,\ch
  \;,
\ee
%,
where $\ch$ is a unit round metric on $S^{n-1}$. Note that
$(b,0)$ are initial data for anti-de Sitter space-time.
%, or a quotient thereof.
We further assume that there exist constants $k\ge 1$, $\alpha>n/2$
and $C>0$ such that
\bel{adsDec}
 |g-b|_b+|\zD g |_b+\cdots+ |\underbrace{\zD\cdots
\zD}_{k \ \mathrm{factors}} g |_b+|K|_b+\cdots+
|\underbrace{\zD\cdots \zD}_{k -1\ \mathrm{factors}} K |_b\le
Ce^{\alpha r}\; . \ee
%
%$$
%
Here $|\cdot|_b$ denotes the norm of a tensor field with respect to
the metric $b$, and $\zD$ is the covariant derivative of $b$.

In particular the definition enforces the vanishing of $\tr_g K$ for
CMC data. Whether or not the data are CMC, \eq{adsDec} implies the
vanishing of the trace-free part of the extrinsic curvature of the
conformal boundary at infinity.% These conditions are not as general
%as one would wish, as they \emph{a priori} exclude large classes of
%smooth hypersurfaces in asymptotically anti-de Sitter space-times;
%we will return to this issue later.

Let $X$ be a Killing vector in the asymptotic region of the
background anti-de Sitter space-time, the  Hamiltonian associated
with the flow along $X$ can be calculated as
follows~\cite{HT,ChAIHP,ChNagyATMP,CJL}: Let $V$ be the normal
component of $X$ with respect to the background adS metric, and let
$Y$ be the tangential component thereof; when defined along a
spacelike hypersurface, such pairs $(V,Y)$ are called KIDs (Killing
Initial Data). Then the Hamiltonian $H(V,Y)$ corresponding to $X$
(which we identify with the couple $(V,Y)$) takes the form:
\be \label{mi}
 H(V,Y):=\lim_{R\to\infty} \frac 1 {16 \pi} \int_{r=R}
\left(\ourU^i(V)+\ourV^i(Y%)\circ\Phi^{-1}\Big
)\right) dS_i
 \;,
\ee
where
\begin{eqnarray} \label{eq:3.3}
 &{}\ourU^i (V):=  2\sqrt{\det g}\;\left(Vg^{i[k} g^{j]l} \zD_j
g_{kl}
+D^{[i}V %(\sqrt{\det g}
g^{j]k} (g_{jk}-b_{jk})\right) \;,
 &
 \\
 &
 {}\ourV^i (Y):=  2\sqrt{\det
g}\;\left(K^i{_j}-K^k{_k}\delta^i_j\right)Y^j \;.\label{eq:3.4}
 &
\end{eqnarray}
Here all indices are space indices, running from $1$ to $n$, and
$\zD$ is the Levi-Civita derivative of the space background metric
$b$.

A preferred set of background Killing vector fields is provided by
those which are $b$-normal to the initial data surface.  The
resulting Hamiltonians are usually interpreted as energies.  In
contradistinction with the asymptotically flat case, where only one
normal background Killing vector field exists, if one assumes that
conformal infinity has spherical space-like sections, then there are
several normal background Killing vector fields. This implies that
there is not a \emph{single} energy, but rather an \emph{energy
functional $M$}. This functional $M$ is uniquely characterised by
$n+1$ numbers $M_\mu$, $\mu=0,1,\ldots,n$, which transform as a
Lorentz vector under asymptotic isometries of $g$,
see~\cite{ChNagyATMP}. (The component $M_0$ coincides with the
Abbott-Deser mass under appropriate restrictions~\cite{ChNagyATMP}.)
It follows that the Lorentzian length of $M_\mu$ is a geometric
invariant of $(\hyp,g)$.  The asymptotically-adS-positive-energy
theorem implies that $M_\mu$ is causal, future
pointing~\cite{GHWSpires,GHHP,Maerten}
(compare~\cite{ChHerzlich,Wang,Zhang:hpet,HertogHollands}), unless
$(\hyp,g,K)$ are initial data for anti-de Sitter space-time. Let us
assume that we are not in this last situation.

 It is convenient to view the
hyperbolic space as a unit spacelike hyperboloid in $\R^{n+1}$, the
latter equipped with the Minkowski metric. Assuming that $M_\mu$ is
timelike,%
\footnote{One expects that $M_\mu$ cannot be null, see~\cite{CMT}
for some partial results.}
after applying an asymptotic isometry to
obtain
$$
 M_\mu=(m,0,\cdots,0)
 \;,
$$
%,
the background Killing vector fields tangent to $\hyp$ can now be
split into rotations and ``boosts". It is customary to define the
rest-frame angular momentum as
$$ j_i:=H(0,\beta_{(i)})\;,$$ where the $\beta_{(i)}$'s are the
generators of rotations of $S^{n-1}$, when embedded in $\R^n$; for
example, in space-dimension $n=3$ a natural choice is
$$\beta_{(i)} = \epsilon_{ijk}x^j\partial_k\;.$$ The numerical values
of the remaining $n$ Hamiltonians generating boost transformations
will be denoted by $c_i$. For initial data which are asymptotically
flat in spacelike directions,  the $c_i$'s have the interpretation
of the centre of mass, and we will retain the name of centre of mass
for the vector $\vec c=(c_i)$.

For reasons   which  are discussed in Section~\ref{Shd} below, from
now on we restrict our attention to $n=3$.   Assuming that $(\hyp,g
)$ is complete, that the dominant energy condition holds,
\bel{pec}
 |\mu|_g \le \rho
  \;,
\ee
where $\mu$ and $\rho$ are as in \eq{c1}-\eq{c2}, and that the total
matter energy as defined by%
\footnote{We take this opportunity to correct~\cite{CMT}, where the
weight factor $e^r$ in \eq{maten} has been inadvertently omitted
from the hypotheses of the positive charges theorem.}
\bel{maten}
 \int_\hyp (1+e^r) \rho\; d\mu_g
\ee
(with $r$ as in \eq{backgmet}) is finite, it is shown
in~\cite{Maerten} (compare~\cite{CMT}) that the positive energy
theorem implies the following inequality
\bel{Jin2}m\ge \sqrt{-\Lambda/3}\sqrt{|\vec c|^2+|\vec j|^2+2|\vec c
\times \vec j|}
 \;,
 \ee
where $\vec c \times \vec j$ is the vector product, while $|\vec j|=
\sqrt{j_1^2+j_2^2+j_3^2}$, etc.

 The inequality also holds if $\hyp$
is complete with boundary, as long as the boundary satisfies one of
the  ``trapping" conditions:  the boundary is either \emph{weakly
future trapped}, which means that
\bel{wft} \tr_h \lambda + h^{ab}K_{ab}\le 0\;, \ee
or \emph{weakly past trapped}, which corresponds to changing the
sign in front of the $K$ term in \eq{wft}. Yet another such
condition
 is obtained~\cite{ChHerzlich,Maerten} by setting
 $k(\nu)=K_{ia}\nu^idx^a$, where the $x^a$'s are coordinates on $\partial \hyp$,
 then the positivity of the global charges will hold if
 \bel{traps}
\tr_h\lambda +|k(\nu)|_h \le  \sqrt{\frac{-2(n-1)\Lambda}{n}}
 \ee
 (see~\cite[Remark~4.8]{ChHerzlich} for a discussion of \eq{traps} when $k(\nu)=0$).

It has been proved in~\cite{CMT} that equality in \eq{Jin2} holds
only for initial data in anti-de Sitter space-time \emph{provided}
the associated space-time has a  Scri with a sufficiently large time
extent. Our application of \eq{Jin2} in Section~\ref{Sami} makes it
clear that it would be of interest to obtain a proof without such a
condition.

\section{Hamiltonian global charges in space-times asymptotically
flat at $\scrip$}
 \label{Sscri}

In this section we briefly review the space-time version of the
approach in~\cite{CJK}.  Let $(\mcM,\fourg)$ be a four-dimensional
space-time with a smooth, or polyhomogeneous, conformal boundary
completion at null infinity $\tmcM=\mcM\cup \scrip$ \emph{\`a la
Penrose}. Let $\hyp$ be a smooth spacelike hypersurface in $\tmcM$
which intersects $\scrip$ transversally at a smooth section
$S=\partial\hyp=\hyp\cap\scrip$. Such a section singles out a six
parameter family of Bondi coordinate systems, by the requirement
that in the chosen Bondi coordinates we have $S=\{u=0\}$.  Now,
every such coordinate system defines a flat background metric
$\bmetric$ in a neighborhood of $S$:
\begin{equation}
  \label{backg}
  \bmetric=\bmetric_{\mu\nu} dx^\mu dx^\nu\equiv
  -du^2-2du\,dr+r^2\hst_{AB}dx^A
  dx^B\;.
\end{equation}
 The resulting metrics are
independent of the Bondi coordinate system chosen, within the six
parameter freedom available, as those coordinate systems differ from
each other by a Lorentz transformation. We can thus define a unique
six parameter family of BMS generators which are singled out by the
requirement that they are tangent to $S$, and that they are Killing
vector fields of the background metric $\bmetric$.

Consider, near $S\subset\scrip$, a Bondi-type coordinate system
$(u,x,v^A)$ as above  with $u\in (-\epsilon,\epsilon)$, $x\in
[0,\epsilon)$, for some $\epsilon
>0$, while the $v^A$'s are coordinates on $S^2$. Here the usual Bondi coordinate $r$ is replaced by $1/x$ so
that the space-time metric $\fourg$, when conformally rescaled by
$r^{-2}$, takes the form
\begin{eqnarray}
\label{gBx}  x^2\;\fourg_{\mu\nu}{\rm d}x^\mu{\rm
  d}x^\nu %&\equiv&  \tfourg _{\mu\nu}{\rm d}x^\mu{\rm
%  d}x^\nu
%\nonumber \\
& = &  -\displaystyle Vx^3 \jexp{2\beta}{\rm d}u^2
+2\jexp{2\beta}{\rm d}u{\rm d}x
\nonumber \\
&  & +  h^{\mB}_{AC}({\rm d}x^A -U^A {\rm
  d}u)({\rm d}x^{C} -U^C {\rm d}u)\;, \\  \displaystyle\frac
{\partial(\det h^{\mB}_{AC})} {\partial x} & = &  0\;.
\end{eqnarray}
If the matter fields decay sufficiently fast then, for smooth
conformally rescaled metrics, one  has the following asymptotics
\begin{eqnarray}
h^\mB_{AB} &=& \bh_{AB}
%\left(1+\frac
%1{4r^2}\chi^{CD}\chi_{CD}\right)
+ \frac{\chi_{AB}(v)}{r} + O(r^{-2})\;, \label{G.1a}
 \\ \beta &=& -\frac{\bh^{AB}
\bh^{CD}\chi_{AC}\chi_{BD}}{32r^2} + O(r^{-3})
 \;, \label{G.1.b0}
 %\nonumber
 \\  U^A &=&
 -\frac{{\ccalD}_B
\chi^{AB}}{2r^2} + \frac{2N^A(v)}{r^3}+
\frac{{\ccalD}{^A}\left(\chi^{CD}\chi_{CD} \right)}{16r^3} \nonumber
+
o(r^{-3})
\, ,
\\
 V &=& r-2\mu_\TBm +
O(r^{-1}) \, , \label{G.1c}
\end{eqnarray}
where $\bh$ is the unit round metric on $S^2$,  ${\ccalD}$ the
corresponding derivative operator, while $\mu_\TBm$ is   the Bondi
mass aspect function.

 In terms of these variables, the
Hamiltonian associated to rotations and boosts
reads~\cite[Eq.~(6.117)]{CJK}
\begin{eqnarray}
  H_L (X,\hyp) &=& -\frac1{64\pi}\int_{S^2}\Big(24 N_A +
  2{\chi}_{AB}{\chi}^{BC}{_{||C}}
    \nn \\ && \phantom{xxxxxxxxx} +
    \frac12(\chi_{BC} \chi^{BC})_{||A}
\Big)X^A\big|_{x=0}\sin\theta\rd\theta
  \rd\varphi \;.\label{newH2}
  \end{eqnarray}
where the vector fields $X$ in \eq{newH2}   belong to the  six
dimensional vector space of $b$-Killing vectors uniquely singled out
by $S=\partial \hyp$.

The above definition has several good properties, discussed
in~\cite{CJK}, some of which are used in Section~\ref{Sobs} below.
For a discussion of alternative definitions of angular momentum at
$\scri$, see~\cite{SzabadosLR}.

\section{The global charges of hyperboloidal initial data sets with $\Lambda=0$}

We continue  with a review of the initial data version of the
analysis in~\cite{CJK}. Consider an asymptotically CMC hyperboloidal
initial data set $(\hyp,g,K)$. In~\cite[Appendix~C.3]{CJL} a
construction is given of an embedding $\iota: \hyp\to \mcM^{\mB}$ of
such an initial data set into a space-time $(\mcM^{\mB},g^{\mB})$
coordinatised as in \eq{gBx}, with the property that the conformal
boundary of $\hyp$ is mapped to $u=0$. Both the embedding $\iota$
and $(\mcM^{\mB},g^{\mB})$ are constructed so that $\iota^* g^{\mB}$
is asymptotic \emph{to infinite order} to $g$ at the conformal
boundary of $\hyp$; similarly the pull-back to $\hyp$ of the
extrinsic curvature of $\iota(\hyp)$ is asymptotic \emph{to infinite
order} to $K$.   The angular momentum and the centre of mass of
$(\hyp,g,K)$ are then defined using \eq{newH2}.

The coordinates $(x,v^A)$ on $\mcM^\mB$, when composed with $\iota$,
induce coordinates  near the conformal boundary of $\hyp$ which will
be denoted by the same symbols.
One can then write $\iota(\hyp)$ as a graph:
$$
 u = \alpha(x, v^A)
 \;, \qquad \alpha(0,v^A)=0
 \;,
$$
and we have (see~\cite{CJL})
\be\label{deralphx}\alpha_{,x}\Big|_{x=0}= \frac{9}{2(\tr_g
K)^2}\;,\ee \be\label{deralphxx}\alpha_{xx}\Big|_{x=0}=-\frac 12
\left({3\over \tr_g  K}\right)^3(\tr_g K)_{,x}
 \;,
 \ee
\be
  \tmet  =  (2\frac{\partial \alpha}{\partial x}+O(x)){\rm d}x^2
+O(x){\rm d}x{\rm d}x^A + (\hst_{AB} + x \chi_{AB} + O(x^2)){\rm
d}x^A{\rm d}x^B\; .
  \label{Ap2}
\ee
Thus the extrinsic curvature of the conformal boundary at infinity,
say $\tlambda_{AB}$, is proportional to $\chi_{AB}$:
\begin{eqnarray}
  \tlambda_{AB} % = - \frac 2 a  \chi_{AB}
  =
-\frac 6 {\tr_g  K}  \chi_{AB}
%= -\frac 12{|d\varomega|_{\tmet}} \chi_{AB}
\; .
  \label{continuity}
\end{eqnarray}
Hence $\tlambda$ vanishes if and only if  $\chi$ does; this will be
relevant shortly.

\section{The angular momentum inequality}
 \label{Sami}

With these preliminaries, we may now state the inequality. Consider
a \emph{CMC} hyperboloidal initial data set $(\hyp,g,K)$ with  $
\dim \hyp=3$, $\tr_gK=\kappa$  and $\Lambda=0$. Suppose that
$(\hyp,g )$ is complete and that the dominant energy condition
\eq{pec} holds. In this section we will assume that
\bean
 && \mbox{\emph{the trace-free part of the extrinsic curvature}}
 \\
 && \mbox{\emph{of the conformal boundary at infinity vanishes};}
\eeal{badcond}
%$$
%
an argument indicating that \eq{badcond} can  be removed will be
presented in  Section~\ref{Sdp} below. (Note, however, that
\eq{badcond} has been invoked in the literature in the context of
CMC hyperboloidal
surfaces~\cite{EardleySmarr,AnderssonIriondo,Goddard}.)  It follows
from \eq{continuity} that this is equivalent to the hypothesis that,
in Newman-Penrose terminology, the associated Bondi cone is
asymptotically shear free. Performing the transformation
\eq{Ktransf}, the initial data set $(\hyp,g,\hat K)$ satisfies the
constraint equations with
$$
 \Lambda = -\frac   {\kappa^2} 3
 \;.
$$
We need to analyse what happens with the global charges under
\eq{Ktransf}. First, using the formulae in \cite[Appendix~F]{CJL}
one checks that, both for translations and rotations,  any trace
terms in \eq{mi} integrate out to zero, so that the extrinsic
curvature contributions to \eq{mi} from $K_{ij}$ and $\hat K_{ij}$
coincide. The same is true for boost generators if \eq{badcond} is
assumed. Next, it follows from \cite[Appendix~C.3]{CJK} that
\cite[Equation~(3.13)]{CJL} holds, which implies that the functional
\cite[Equation~(3.11)]{CJL} coincides with \eq{mi} (see
\cite[Equation~(3.14)]{CJL}). Letting $ m$ be the Hamiltonian mass
of $(\hyp,g,\hat K)$, and  $m_\TBm$  the Trautman-Bondi mass of
$(\hyp,g,K)$, the equality
\bel{masseq}
 m = m_\TBm
 \;.
\ee
%$$
%
follows now from Theorem~5.3 of \cite{CJL}.

For the remaining charges, observe that \emph{under \eq{badcond}}
the integrals \eq{mi} are equal to their linearisations.  Now, it
has
been shown in \cite[Appendix~B]{CJL} that,%
\footnote{Note that the terms quadratic in $\chi$ in the last
equation of \cite[Appendix~B]{CJL} might seem to be incompatible
with the fact that a linearised expression is considered. This
apparent contradiction is resolved by observing that some
coefficients of the metric, which enter linearly in the integral,
are themselves quadratic  in the free Bondi functions $\chi$ and
their derivatives. }
again under \eq{badcond}, the linearisation of the functional
\cite[Equation~(3.14)]{CJL} equals the linearisation of the
functional $H_{\mathrm{boundary}}$ of \cite{CJK}. The calculations
in~\cite[Sections~6.4 and 6.5]{CJK} then show that the angular
momenta of $K$ and $\hat K$ coincide. Now, the centre of mass  for
$(\hyp,g,\hat K)$ is calculated using only the first term at the
right-hand-side of~\cite[Eq.~(6.57)]{CJK}, while the calculation for
$(\hyp,g, K)$ uses the whole right-hand-side of that equation.
Nevertheless, both quantities are equal under \eq{badcond}.

If we furthermore assume that $\rho$ decays fast enough so that the
total energy as defined by \eq{maten} is finite, then all the
conditions needed for \eq{Jin2} are met, and we conclude that
\bel{Jin2n}m_{\TBm}\ge\frac{ |\tr _gK|} 3\sqrt{|\vec c|^2+|\vec
j|^2+2|\vec c \times \vec j|}
 \;.
 \ee
Here $m_{\TBm}$ is the Trautman-Bondi mass, $\vec j$ is the angular
momentum vector (the Hamiltonian associated with rotations) in the
rest frame (i.e., a conformal frame in which  space-momentum
vanishes), and $\vec c$ is the centre of mass (the Hamiltonian
associated with boosts) in that frame. In particular we have the
striking   bounds
\bel{Jin3}m_{\TBm}\ge\frac{ |\tr _gK|} 3|\vec j|\;,\quad
m_{\TBm}\ge\frac{ |\tr _gK|} 3|\vec c|\;.\ee
In the light of the earlier discussion of (2.8), it is expected that
equality in \eq{Jin2n} can occur only for initial data in Minkowski
space-time; it would be of interest to prove this.

\section{A possible direct proof}
 \label{Sdp}

In this section we indicate an argument that could remove the
restrictive condition \eq{badcond}. We start with some notation. In
space-time dimension $n$, we view the hyperbolic space as the open
unit ball $B^n(1)\subset \R^n$ equipped with the metric $b=\backn =
\omega^{-2}\delta$, where $\delta$ is the standard flat metric on
$\R^n$, and
$$
 \omega= \frac{1-|x|^2}2\;.
$$
If we write the Minkowski metric $\eta$ as $-dt^2+\delta_{ij}dy^i
dy^j$, and  set
\bel{mincoorhyp}
 \tau = t- \sqrt{1+|y|^2}\;,\quad y^i = \omega^{-1} x^i \;,\quad r=|x|\;,
\ee
we obtain
$$
\eta = -d\tau^2 + \omega^{-2}(-2rd\tau dr + \delta_{ij}dx^i dx^j)
 \;.
$$
The KID-decompositions of the Minkowskian Killing vectors at
$\hyp:=\{\tau=0\}$ read
\beaa
 &
 %X_{(0)}=
 % \partial_t=  \underbrace{\frac{1+|x|^2}{1-|x|^2}}_ {V_{(0)}}n  \underbrace{- x^i \partial_{x^i}}_{Y_{(0)}}\;,
 \partial_t=V_{(0)}n+Y_{(0)}= \frac{1+|x|^2}{1-|x|^2} n - x^i \partial_{x^i}\;,
 &
 \\
 &
 %X_{(i)}=
 \partial_{y^i}=V_{(i)}n+Y_{(i)}
 = -\omega^{-1} x^i n + \omega\partial_{x^i} + x^i x^j \partial_{x^j}\;,
 &
 \\
 &
 %X_{(i)}=
 t\partial_{y^i}+y^i \partial_t=0\cdot n+C_{(i)}=\frac{1+|x|^2}2\partial_{x^i} - x^i x^j \partial_{x^j}
 \;,
 &
 \\
 &
 y^i \partial_{y^j}-
 y^j \partial_{y^i}=0\cdot n+ \Omega_{(i)(j)}= x^i \partial_{x^j}-
 x^j \partial_{x^i}
 \;,
 \eeaa
 %$, and letting
where $n$ is the unit normal to $\hyp$.

The standard proof of positivity of \TB mass proceeds by solving the
Witten equation:
\bel{Wittens}
  \gamma^i \nabla_i \psi =0\;, \ \mbox{ where } \
 \nabla_i:=D_i+ \frac 12 K_{ij}\gamma^j \gamma^0 \;.
\ee
%$$
%
One further requires $\psi$ to asymptote  to spinors $\zpsi$ which
are restrictions to a hyperboloid of covariantly constant spinors in
Minkowski space-time. For hyperboloids with $ \zK_{ij}=-b_{ij}$ the
spinors solve%
\footnote{We use the conventions of~\cite{CJL}, in which the
standard unit future hyperboloid in Minkowski space-time $\R^{1,n}$
satisfies $\tr_gK=-n$.}
\bel{parallel}
 \zD_i \zpsi = \frac 12 \gamma_i \gamma^0 \zpsi
 \;.
\ee
In the obvious spin frame associated with the above conformal
representation%
\footnote{More precisely, we take a spin frame which projects to the
frame $\theta^i=\omega^{-1}dx^i$, with $e_i$ dual to $\theta^i$, and
a local basis of the spinor bundle in which the $\gamma^\mu$'s are
constant matrices.},
the solutions of \eq{parallel} read
\bel{immKs}
 \psi_u = \omega^{-1/2}(1+ x^k \gamma^k\gamma^0)u %\;,
 \ee
(summation over $k$), where $u$ is a spinor with constant entries,
while the anti-Hermitian matrices $\gamma^k$ with constant entries
satisfy the flat space Clifford relations
$$
\gamma^i \gamma^j + \gamma^j \gamma^i = -2\delta^{ij}\;.
$$
 (The $\psi_u$'s exhaust the space of
solutions because the map which assigns $u$ to, e.g., $\psi_u(0)$ is
a bijection). Further, $\gamma^0$ is a Hermitian matrix, with
constant entries, satisfying
$$
(\gamma^0)^2=1\;,\qquad \gamma^0 \gamma^j + \gamma^j \gamma^0 = 0\;.
$$
(The spinor bundle can always be chosen so that such a matrix
exists.) The KID $(\Nu ,Y^i_u)$ associated to $\psi_u$ takes the
form
\beal{KID1}
 \Nu &:=& \langle \psi_u, \psi_u \rangle = 2\Big(|u|^2
 \underbrace{\frac{1+|x|^2}{1-|x|^2}}_{V_{(0)}} -\langle u, \gamma^k
 \gamma^0 u\rangle
 \underbrace{\frac{(-2)x^k}{1-|x|^2}}_{V_{(k)}}\Big)\;,
 \\
 Y^i_u \partial_i &:=& \langle \psi_u, \gamma^0\gamma^i \psi_u
 \rangle e_i
 \nonumber
 \\
  &=&
 -2\Big( |  u|^2
 \underbrace{x^i\partial_i }_{Y_{(0)}}+\langle u, \gamma^k\gamma^0 u\rangle \underbrace{\Big(\frac{1-|x|^2}2\delta^i_k
 +
 {x^i x^k}\Big)\partial_i}_{Y_{(k)}}\Big)
 \;.
% \nonumber
% \\
% &&
 \eeal{KID2}
This, together with the usual Witten argument, implies that the
boundary term in the Witten equation will only carry information
about the global charges associated with space-time translations of
$\R^{1,n}$.

Now, our argument so far leading to the angular momentum bound can
be rephrased as follows: instead of  \eq{Wittens} one considers
\bel{Wittenmo}
  \gamma^i \hnabla_i \psi =0\;, \ \mbox{ where } \
 \hnabla_i:=D_i+ \frac 12 \Big(K_{ij}-\frac {\tr_gK} n g_{ij}\Big)\gamma^j \gamma^0 -   \frac {i\tr_gK} {2n}\gamma_i
 \;,
\ee
%$$
%
where the $\psi$'s   asymptote now  to imaginary Killing spinors
$\hpsi$ of $b$ which, for $\tr_g K = -n$, solve
\bel{parallelo}
 \zD_i \hpsi = - \frac i2 \gamma_i   \hpsi
 \;.
\ee
Those take the form
\bel{immKsoro}
 \hpsi_u = \omega^{-1/2}(1-i x^k \gamma^k)u %\;,
 \ee
(summation over $k$), where $u$ is again a spinor with constant
entries. Instead of \eq{KID1}-\eq{KID2}, the  KID $(\hat V_u ,\hat
Y^i_u)$ associated to $\hpsi_u$ takes the form
\beal{KID1o}
 \hat V_u &:=& \langle \hpsi_u, \hpsi_u \rangle = 2\Big(|u|^2
 \underbrace{\frac{1+|x|^2}{1-|x|^2}}_{V_{(0)}} +\langle u, i\gamma^k
 u\rangle
 \underbrace{\frac{(-2)x^k}{1-|x|^2}}_{V_{(k)}}\Big)\;,
 \\
 \hat Y^i_u \partial_i &:=& \langle \hpsi_u, \gamma^0\gamma^i \hpsi_u
 \rangle e_i
 \nonumber
 \\
  &=&
 2\langle u, \gamma^0\gamma^k u\rangle \underbrace{\Big(\frac{1+|x|^2}2\delta^i_k -
 {x^i x^k}\Big)\partial_i}_{C_{(k)}}
 \nonumber
 \\
 &&
 +  \frac 12 \langle u, i\gamma^0(\gamma^k\gamma^i-\gamma^i\gamma^k) u\rangle
 \underbrace{(x_k\partial_i-x_i\partial_k)}_{\Omega_{(k)(i)}}
 \;,
% \nonumber
% \\
% &&
 \eeal{KID2o}
so that the boundary term in the Witten identity will carry now
information about all global charges.

We are ready to prove that the existence of solutions of
\eq{Wittenmo} with the above boundary condition, without assuming
the vanishing of $\chi$. Indeed, from inspection of the positivity
proof of~\cite[Section~5.4]{CJL}  one infers that one needs to
justify
\bel{Ltwo}
  \gamma^i \hnabla_i \hpsi \in L^2
  \;,
\ee
compare the proof of Lemma~5.9 in~\cite{CJL}.  In what follows
notations and
conventions of% this last reference %
~\cite{CJL}
are used unless explicitly indicated otherwise.%
\footnote{We take the opportunity to point out the following
misprints there: first, $\gamma^0$ is assumed to be hermitian and
$\gamma^i$ -- anti-hermitian, in spite of what is said at the
beginning of page 122 of~\cite{CJL}. Next, $\sqrt{\det g}$ should be
estimated as $O(x^{-3})$ in the penultimate displayed equation of
Appendix D of~\cite{CJL}. In \cite[Equation~(5.14)]{CJL} the factor
$1/4\pi$ should be $4\pi$.}
 Now, after a
constant rescaling so that $\tr_gK=-n$, from \eq{Wittenmo} we obtain
\bean
  \gamma^i \hnabla_i \hpsi  & = &
 \gamma^i D_i\hpsi+ \underbrace{\Big(K_{ij}-\frac {\tr_gK} n g_{ij}\Big)\gamma^i\gamma^j \gamma^0
 }_{0}\hpsi-
  \frac {ni } {2
 }\hpsi
 \;.
\eeal{Wittenmo2}
By \eq{parallelo} we have
\bel{Xhpsi} X(\hpsi) = \frac 14 \zomega_{ij}(X)\gamma^i\gamma^j
\hpsi- \frac i 2 \sum_\ell\mathring X^\ell\gamma^\ell \hpsi
 \;,
\ee
where $\mathring X^\ell$ are the components of $X$ in the
$b$--orthonormal  frame $\repe_i$ as in \cite[Appendix~C]{CJL}: $X=
\mathring X^i \repe_i$. In \eq{Xhpsi} we have indicated explicitly
the summation over $\ell$ since both $\ell$'s are  superscripts
there. Letting $\repf_i=\hat M_i{}^j \repe_j$
be  the $g$-orthonormal frame as in \cite[Appendix~C]{CJL}, %~\cite{CJL},
it follows that
\begin{eqnarray} %\hspace*{-1cm}
\gamma^\ell\hnabla_\ell\hpsi & = & \gamma^\ell\repf_\ell(\hpsi)
-\frac14
\omega_{ij}(\repf_\ell )\gamma^\ell\gamma^i\gamma^j   \hpsi
 - \frac {ni} 2 \hpsi \nonumber \\
& = &   \frac14 \left( \zomega_{ij}(\repf_\ell ) - \omega_{ij}(\hat
\repf_\ell ) \right) \gamma^\ell\gamma^i\gamma^j \hpsi
 - \frac i
2\sum_{j}(\hat M_{i}{}^ { j}-\delta_{i}^{j})\gamma^i\gamma^j \hpsi
% - \frac{n i} 2
%\hpsi
 \; .
  \phantom{xxx}\label{c1rs}
\end{eqnarray}
It has been shown in~\cite[Appendix~D]{CJL} that the first term in
\eq{c1rs} can be estimated by $C x^2 |\hpsi|$, which in turn implies
that it is in $L^2$. Next, by \cite[Equations (C.21), (C.22), (C.40)
and (C.47)]{CJL}, both the anti-symmetric part and the trace of
$\hat M_{i}{}^ { j}-\delta_{i}^{j}$  are $O(x^2)$, and \eq{Ltwo}
follows.  This, together with the arguments in~\cite{CJL,ChHerzlich}
proves existence of the relevant solutions of the Witten equation.
In retrospect, the calculation here is shorter than the one for the
original positivity proof, albeit  applying to CMC initial data
only.

To complete the proof of \eq{Jin2n} without the restrictive
condition \eq{badcond} one needs to analyse the boundary term that
appears in the Witten identity associated to the operator
\eq{Wittenmo}. We are planning to return to this in the near future.

\section{The conformal method} \label{sScm}

Given a space-time $(\mcM,g)$, it is far from clear whether or not
$\mcM$ contains \emph{any} complete CMC surfaces (see, however,
\cite{AnderssonIriondo}). Furthermore, it is not clear whether or
not those surfaces will be sufficiently differentiable at $\scrip$
as needed above. Therefore it is reasonable to raise the question of
the range of applicability of our bounds. Recall, now, that the
conformal method provides a construction of all, say vacuum, CMC
general relativistic initial data sets. In the hyperboloidal context
one prescribes a non-zero value of $\tr_gK$, as well as an arbitrary
conformally compactifiable Riemannian manifold $(\hyp, \mathring g)$
equipped with a seed symmetric trace-free tensor, say $A$, and
constructs $(\hyp,g,K)$ by solving a set of elliptic equations,
see~\cite{AndChDiss} and references
therein. In such a construction the resulting initial data set will
satisfy condition \eq{badcond} if and only if the trace-free part of
the extrinsic curvature of the conformal boundary at infinity of
$\mathring g$ vanishes.
Since $\mathring g$ and $A$ can be chosen arbitrarily, subject to a
finite number of compatibility conditions at the conformal
boundary~\cite{PRLetter}, we conclude that there exists an infinite
dimensional family of vacuum initial data sets for which \eq{Jin2n}
provides  a non-trivial upper bound for $\vec j$ and $\vec c$ in
terms of the total mass. The associated globally hyperbolic vacuum
developments~\cite{Friedrich:tuebingen} provide, in turn, examples
of space-times containing hypersurfaces satisfying the hypotheses of
our inequality.

\section{Obstructions to existence of CMC surfaces}
 \label{Sobs}
Note that \eq{Ap2} shows that $\tilde \lambda_{AB}$ is the same for
all CMC surfaces asymptotic to a given cut of $\scri$. This leads to
 the following unexpected consequence of our analysis: whenever $|\vec j|+|\vec c|\ne
0$ there exists an upper bound on $|\tr_g K|$ for complete
hyperboloidal CMC surfaces satisfying%
\footnote{In view of the analysis of Section~\ref{Sdp}, it is rather
likely  that \eq{badcond} is not needed for the discussion of this
section.}
\eq{badcond} (without boundary, or with boundaries on or beyond
horizons) which asymptote to smooth cuts $S$ of $\scri$, namely
\bel{Jin2nud} |\tr _gK|\le\frac{3m_{\TBm}}{ \sqrt{|\vec c|^2+|\vec
j|^2+2|\vec c \times \vec j|}}
 \;.
 \ee

\subsection{CMC surfaces in Schwarzschild}

\Eq{Jin2nud} does not lead to any restrictions on $\tr_g K$ for CMC
hypersurfaces in Schwarzschild space-time which asymptote to
spherically symmetric cuts of $\scrip$, and indeed there are
none~\cite{MalecOM2003}. Consider, however, cuts $S_\alpha$ of the
Schwarzschildian $\scrip$ which are obtained by applying a
translation $u\to u+\alpha$ to $S_0=\{u=0\}$, where $\alpha$ is a
linear combination of $\ell=0$ and $\ell=1$ spherical harmonics. As
shown in~\cite[Section~6.6]{CJL}, all such cuts have vanishing
angular momentum. More generally, it is shown
in~\cite[Section~6.7]{CJL} that for all stationary space-times the
Hamiltonian angular momentum is independent of the cut of $\scrip$
chosen, so the discussion that follows applies to any stationary
space-time with matter satisfying the dominant energy condition. It
is also shown in~\cite[Sections~6.6 and 6.7]{CJL} that the change of
centre of mass of $S_\alpha$ can be calculated using the standard
special-relativistic rule: under a translation by a vector $\vec a$
orthogonal to the momentum the centre of mass is shifted by $m \vec
a$. Since \eq{badcond} is preserved under translations, from
\eq{Jin2nud} we conclude that for any translation  $\vec a$,
\emph{the associated cut $S_\alpha$ in the Schwarzschild space-time
cannot span a complete CMC surface meeting $\scrip$ smoothly (or
$C^2$ and polyhomogeneously) with }
\bel{Kveca}
 |\tr_g K|> \frac 3 {|\vec a|}
 \;.
\ee
An identical conclusion is reached in space-times which are
stationary near $\scrip$ and have zero angular momentum, and a
similar conclusion without assuming that $\vec j =0$.

\subsection{CMC hypersurfaces in Kerr space-time?}
 \label{sSKerr}

Both the tensor field $\chi$, and the centre of mass vanish for the
family of $\{u=\const\}$ cuts of $\Scrip$ in Kerr space-time, where
$u$ is an outgoing Eddington-Finkelstein coordinate,  and for these
we obtain
\bel{Kerrbound}
 |\tr_gK|\leq\frac 3{|a|}\;,
\ee
where $a$ is the usual angular momentum parameter in the Kerr
metric, for any complete CMC surface spanned by those cuts. As
above, it follows immediately that no such surfaces exceeding this
bound exist.

We wish to present an argument which suggests strongly that no such
hypersurfaces exist in Kerr at all. Suppose, for contradiction, that
there exists a complete spacelike hypersurface $\hypkz$ in Kerr
space-time, satisfying \eq{badcond}, with $\tr_g K=\kappa_0$, for
some $\kappa_0<0$, with two spherical boundaries lying on two
different components of $\scri^+$. We further assume that $\hypkz$
is contained within four diamond-shaped blocks  of the usual maximal
analytic extension of Kerr, on two of which  $r_-<r<r_+$, while
$r>r_+$ on the remaining, asymptotically flat, ones.   Choose any
$\kappa$ more negative than $-3/|a|$ (and thus smaller than
$\kappa_0$) and let $\hyp_n$ be a sequence of CMC surfaces with
$\tr_g K=\kappa$ such that the boundary of $\hyp_n$ consists of two
spherical components lying on $\hypkz$, with $\partial \hyp_n$
approaching $\scri$ as $n$ tends to infinity. Such $\hyp_n$ exist by
the results in~\cite{Bartnik84,Bartnik:variational}, because
$\hypkz$ provides an upper barrier, while a lower barrier is
provided by the boundary of the past domain of dependence, say
$\mcD_n^-$, of that subregion of $\hypkz$ which is bounded by
$\partial \hyp_n$. To see that $\mcD_n^-$ is conditionally compact,
note that it  must be included in the region which is delimited to
the future by $\hypkz$, and which
is delimited to the past%
\footnote{This follows from the fact that in Eddington-Finkelstein
coordinates one has $g_{rr}=\Gamma^\mu_{rr}=0$, so that the curves
$u=u_0$, $\theta=\theta_0$, $\varphi = \varphi_0$ are null
geodesics.}
 by the hypersurfaces $u=u_0$, and $\hat
u=u_0$, where $u$ is an Eddington-Finkelstein retarded coordinate
the level-sets of which provide cuts of $\scrip$ in the first
asymptotic region, while $\hat u$ is the analogous
Eddington-Finkelstein coordinate asociated to the second
asymptotically flat region, with $u_0=\min(\inf_{\partial \hypkz}u,
\inf_{\partial \hypkz}\hat u)$. (Note that one of $u$ and $\hat u$
is actually an advanced Eddington-Finkelstein coordinate $v$ in the
relevant region $r_-<r<r_+$.) This proves that the compactness
condition needed for Bartnik's theorem~\cite{Bartnik:variational} of
existence of smooth solutions of the Dirichlet problem is satisfied.
(An alternative height bound to the past is obtained by the level
sets of $r$  near $r_-$, which are crushing~\cite{EardleySmarr} as
$r\to r_-$.) By Bartnik's interior estimates the sequence $\hyp_n$
converges, in the compact-open topology, to some smooth hypersurface
$\hyp_\kappa$. If one could show
 --- which isn't clear
(compare~\cite{AnderssonIriondo} where the conditions for the
construction of barriers near the boundary preclude a non-vanishing
$J$)
--- that $\hyp_n$ is uniformly spacelike in the conformally rescaled space-time, with a bound independent of $n$, one would
obtain a smooth spacelike CMC surface $\hyp_\kappa$ spanned on
$\partial \hypkz\subset \scrip$. If one could further show
--- which is likely, using the results in~\cite{AndChDiss} --- that
$\hyp_\kappa$ is smooth at $\scri$ (polyhomogeneous and $C^2$ would
suffice~\cite{CJL}; compare~\cite{Stumbles}), one would obtain a
contradiction with \eq{Kerrbound} for $\kappa$ large negative. It
would then follow that no CMC hypersurfaces $\hypkz$ as assumed
above exist in Kerr.

\section{Higher dimensions}
\label{Shd}

It is interesting to enquire what happens in higher dimensions.
Indeed, the positive charges theorem has been proved for
hyperboloidal initial data with $\Lambda<0$ with a spherical
conformal infinity under the assumption that $\hyp$ is spin
(compare, however,~\cite{AnderssonGallowayCai}), together with the
asymptotic conditions~\eq{adsDec}~\cite{CMT}; note that those
require the vanishing, up to an overall conformal factor, of
$\lfloor n/2 \rfloor$ derivatives of the conformally rescaled metric
at the conformal boundary at infinity. Assuming the latter
condition, we expect the transformation \eq{Ktransf} to map all the
global charges at null infinity to the adS ones, but no such
analysis has been carried out so far. Now, an easy way out is to
\emph{define} the charges at null infinity as the values of the adS
ones after the transformation \eq{Ktransf} has been performed. Under
suitable global hypotheses, this gives immediately the global
charges inequalities of~\cite{CMT}%
\footnote{See Theorem~2 of the published version, which is
Theorem~3.1 of the arxiv version.}
in any dimension  $n\ge 3$. It is then unfortunate that no explicit
sharp inequalities are known in space-time dimensions higher than
seven. In any case, it would be preferable to express the
inequalities in terms of global charges directly definable at
$\scrip$, compare~\cite{HollandsIshibashi}. Furthermore, similarly
to $n=3$, we expect the asymptotic conditions \eq{adsDec} to be
overly restrictive for a proper understanding of null infinity.

\bigskip

\noindent{\sc Acknowledgements} We  thank Robert Bartnik for useful
comments.

\bibliographystyle{amsplain}
%\bibliographystyle{/usr/share/texmf/tex/revtex/prsty}
%\bibliography{%ptjjk,
%../../references/newbiblio,%
%../../references/reffile,%
%../../references/bibl,%
%../../references/Energy,%
%../../references/hip_bib,%
%../../references/netbiblio}
\bibliography{%ptjjk,
../references/newbiblio,%
../references/newbib,%
../references/reffile,%
../references/bibl,%
../references/Energy,%
../references/hip_bib,%
../references/netbiblio,../references/addon}
%\texttt{\input{READMEl}}

\end{document}